\documentclass[preprint,12pt]{elsarticle}
\usepackage{amssymb}
\usepackage[nodots,nocompress]{numcompress}

\biboptions{super}
\journal{Physical Review A}
\input{tcilatex}
\begin{document}

\begin{frontmatter}


\title{The mechanism of anisotropic\ exchange interaction in superconducting
iron arsenides}


\author[1]{Xiangang Wan\corref{cor}}
\cortext[cor]{Tel.: +86-25-83595330;fax:+86-25-83595535
}
\ead{xgwan@nju.edu.cn}

\author[1]{Jinming Dong}

\author[2]{Xi Dai}
\address[1]{National Laboratory of Solid State Microstructures and Department of
Physics, Nanjing University, Nanjing 210093, China}
\address[2]{Beijing National Laboratory for Condensed Matter Physics, and
Institute of Physics, Chinese Academy of Sciences, Beijing 100190, China}
\begin{abstract}
Using a combination of linear response theory and constrained orbital
hybridization approach, we study the mechanism of magnetic exchange
interaction of iron-based superconductor. We reproduce the observed highly
anisotropic exchange interaction, and our constrain-orbital calculation
unambiguously identifies that the anisotropic feature of exchange
interaction is not sensitive to the unequal d$_{xz}$/d$_{yz}$ orbital
population.

\end{abstract}


\end{frontmatter}


\label{1} 




The discovery of high-temperature superconductivity in iron arsenides has
attracted intense research interests\cite%
{Review-Nat-Phys,FeAs-1,FeAs-2,FeAs-3,FeAs-4}. While the mediator of pairing
in these systems remains officially unidentified, a large amount of
circumstantial evidence points to magnetic spin fluctuations. Therefore,
tremendous amount of efforts have been devoted to understand the magnetic
properties\cite{FeAs-4,Nest-1,Nest-2,neutron ani J,Yildirim J,Z.Y. Lu,Q.M.
Shi,Neutron J,Nematic,M.J.Han J,Johannes Mazin,Wei Ku OO}.

However, despite vast efforts, the nature of magnetism in the iron-based
superconductor is still a hotly debated topic\cite{Review-Nat-Phys}. Early
theoretical studies suggest that superconducting iron arsenides have an
antiferromagnetic spin-density-wave (SDW) instability due to Fermi-surface
nesting\cite{Nest-1,Nest-2}. Neutron scattering experiment\cite{neutron ani
J} confirms that LaFeAsO indeed exhibits the predicted stripe
antiferromagnetic (S-AFM) long-range ordering followed by a small structural
distortion. However, the observed magnetic moment is much small than the
theoretical one\cite{neutron ani J}. Moreover, although the general picture
fits with a SDW model, there remain problems of matching to a purely
itinerant scenario. In particular, the increased conductivity found in SDW
state is not expected if a portion of the carriers become gapped\cite%
{Review-Nat-Phys}. Alternatively, a Heisenberg magnetic exchange model had
been proposed to explain the magnetic behavior\cite{Yildirim J,Z.Y. Lu,Q.M.
Shi}. It had been suggested that nearest-neighbor and next-nearest-neighbor
interactions between local Fe moments are both antiferromagnetic and of
comparable strength\cite{Yildirim J,Z.Y. Lu}, which results in a magnetic
frustration. These frustrating effects have been used to explain the
structural phase transition and small ordered moment\cite{Yildirim J}. It
was also suggested that the structural transition is actually a transition
to a "nematic" ordered phase which will occur at a higher temperature than
the SDW transition\cite{Nematic}.

On the other hand, a short-range and highly anisotropic exchange interaction
had been predicted theoretically\cite{M.J.Han J} and confirmed by the
neutron scattering measurement subsequently\cite{neutron ani J}. To
understand this unexpected anisotropy is a hot topic \cite{Wei Ku OO,Model
OO,SO model,3 band model Dagotto}. As a natural way to break the symmetry,
orbital ordering (OO) had attracted intensive research attention\cite{Wei Ku
OO,Model OO,SO model,3 band model Dagotto}, and there is increasing
experimental evidence about the orbital physics\cite{ARPES dzx}. Band
structure\ calculation proposes that the degeneracy between d$_{xz}$ and d$%
_{yz}$ orbital had been lifted and there is a ferro-orbital ordering, which
results in not only the strong anisotropic exchange but also structural
transition\cite{Wei Ku OO}. However, the electronegativity of As is much
smaller than that of O, the crystal-field effect upon the 3\textit{d}
orbitals of Fe is much weaker than in transition metal oxides, consequently
the orbital polarization is quite small. The OO had also been supported by
the model calculation, but it is not clear whether the exchange anisotropy
is related to OO or not\cite{Model OO}. Therefore, a extensive study about
the mechanism of exchange interaction is an important problem. In this work
we address this issue using the linear response approximation\cite{linear
response for J} as well as a recently developed constrained orbital
hybridization approach\cite{Constrained-hybridiztion-scheme}. While our
linear response approximation reproduce the known anisotropic exchange
interaction, our constrained orbital calculation allows us to provide
theoretically a conclusive insights to various contributions to magnetic
exchange interactions.

We perform our electronic structure calculations based on the
full-potential, all-electron linearized-muffn-tin-orbital (LMTO) method\cite%
{FP-LMTO}. Since for this system local spin density approximation (LSDA) can
give reasonable results\cite{Not U,LSDA is OK}, we therefore adopt it as the
exchange-correlation potential. With the electronic structure information,
we estimate the exchange interaction \textit{J} based on a magnetic force
theorem\cite{force theorem} that evaluates linear response due to rotation
of magnetic moments\cite{linear response for J}. This technique has been
used successfully for evaluating magnetic interactions in a series of
compounds\cite{M.J.Han J,linear response for J,HTC J,EuX}. The main results
and conclusions are found to be the same for all iron arsenides, we
therefore focus on LaFeAsO at the following.

\begin{figure}[tbp]
\includegraphics[width=3.5in]{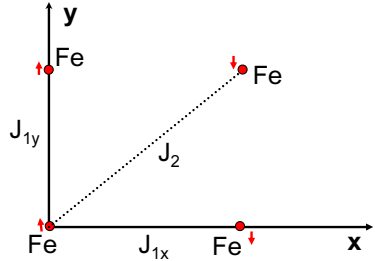}
\caption{Definitions of \textit{x}, \textit{y} axis, the nearest neighbor
exchange interaction $J_{1x}$, $J_{1y}$ \ and the next nearest neighbor
exchange interaction $J_{2}$.}
\label{fig1}
\end{figure}

The calculations are performed on the high-temperature tetragonal structure%
\cite{Low temp str}. The \textit{x} and \textit{y} axes are taken to be
along the Fe-Fe bond direction, with the \textit{x} axis chosen along the
AFM ordered direction of S-AFM as shown in Fig.1. Our calculated ground
state properties, including the magnetic ordering configuration, density of
state and band structure, are found to be in good agreement with previous
theoretical results\cite{M.J.Han J}. Based on the electronic structure
information, we evaluate the interatomic exchange constants as an integral
over the q space using (8,8,8) reciprocal lattice grid. Our numerical
results show that despite the metallic nature, the exchange interaction is a
short range one with the magnetic coupling further than the second nearest
neighbor to be almost equal to zero. The short-range feature of the exchange
interaction may be caused by the small density of state at Fermi energy. We
reproduce the experimental observed strong anisotropic near-neighbor
exchange interaction\cite{neutron ani J}. With the definition of positive 
\textit{J} meaning the antiferromagnetic coupling, our numerical data of 
\textit{J}$_{1x}$, \textit{J}$_{1y}$ and \textit{J}$_{2}$ are 47.9, -8.0 and
21.0 meV, respectively, which are in good agreement with the previous
theoretical results\cite{M.J.Han J}.

Our LSDA calculation confirms that there is a small orbital polarization,
and the difference between the occupation of d$_{xz}$ and d$_{yz}$ orbital
is 0.135, which is very closed to the previous theoretical work (0.141)\cite%
{Orbital occ}. The magnetic moment at d$_{xz}$ and d$_{yz}$ orbital are
0.202 and 0.361 $\mu _{B}$ respectively, which is also consistent with the
previous calculation (0.149 and 0.338 $\mu _{B}$)\cite{Orbital occ}. After
reproducing the orbital/spin polarization, we made a calculation of \textit{J%
}'s with an artificial constrained external potential applied to the d$_{xz}$
orbital of Fe to adjust its energy consequently to control the orbital
occupation, so that we can check the exact effect of unequal d$_{xz}$/d$_{yz}
$ orbital population\cite{Constrained-hybridiztion-scheme}. As shown in
Fig.2(a), J$_{2}$ almost does not depend on the shifting of \textit{d}$_{xz}$
level and the associated orbital polarization, which is contrary to the
suggestion of strong dependence in Ref.\cite{Wei Ku OO}. Although the value
of J$_{1x}$ and J$_{1y}$ do depend on the OO, but as shown in Fig.2(b) and
Fig.2(c), even the d$_{xy}$ and d$_{xz}$ orbital has the same occupation
(i.e., OO equal to zero), there is still strong anisotropy between them (J$%
_{1x}$ is almost twice larger than J$_{1y}$). Thus, we can conclude that the
d$_{xz}$ and d$_{yz}$ orbital do have unequal population, but the
anisotropic exchange is not related to it. 
\begin{figure}[tbp]
\includegraphics [height=3.5in] {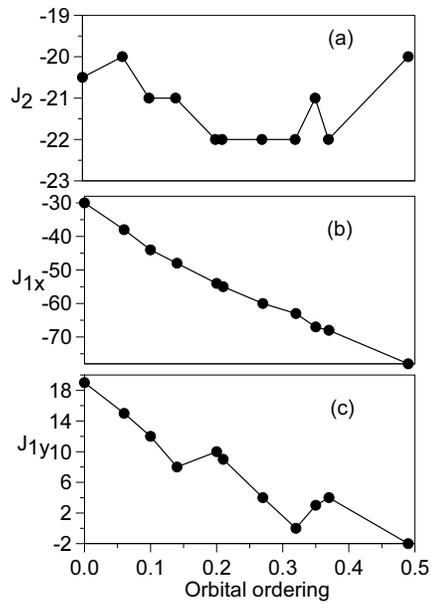}
\caption{The relation between exchange interaction and orbital polarization.
(a) is J$_{2}$; (b) is J$_{1x}$; (c) is J$_{1y}$. The x-axis is orbital
ordering (i.e. the difference between the occupation of d$_{xz}$ and d$_{yz}$
orbital), y-axis is the strength of exchange interaction (in meV).}
\end{figure}

It is well known that the strength of hybridization between two orbitals
strongly depends on their energy difference, therefore the exchange
interaction will be sensitive to the shifting of special orbital if this
orbital participate in the exchange process. We thus perform the
constrained-hybriziation approach\cite{Constrained-hybridiztion-scheme} to
exactly analyze the possible virtual exchange mechanism directly. This
technique has been used successfully in perovskite ruthenates and Europium
Monochalcogenides\cite{Constrained-hybridiztion-scheme,EuX}. It turns out
that a upshift of 5\textit{d} orbital of La or a downshift of 2\textit{p}
orbital of O does not affect the exchange interaction. Therefore, the
exchange process happens almost completely in the FeAs layer, and the
inter-layer exchange interaction is negligible.

In additional to \textit{d}$_{xz}$, we also shift other 3\textit{d} orbitals
of Fe. Shifting the 3$d$ orbitals changes the orbital occupation, however
only shifting d$_{xy}$ orbital has considerable effect on J$_{2}$. Since As
anion is located above the center of the Fe plaquette, one can expect that
the hybridization between Fe-d$_{xy}$ and As-p$_{x\pm y}$ is strong. Thus,
our numerical results clearly show that J$_{2}$ is mainly contributed by the
As-bridged antiferromagnetic superexchange. In contrast to J$_{2}$, all 3$d$
orbitals have large effect on J$_{1x}$ and J$_{1y}$, which indicates the
importance of exchange interaction due to the direct hopping between
nearest-neighbor Fe 3$d$ electron.

It is well known that the interatomic magnetic interaction basically is a
band structure effect, and the spin ordering affects the covalency and
details of the bonding topology. Therefore, it is not surprised that the
exchange interaction depends on the magnetic configuration. For example,\
our additional calculation shows that even for NiO, which has well defined
local moment, there is about 10\% difference between the \textit{J} from
AFM\ and FM\ configuration calculation. The magnetism in iron arsenides is
much more itinerant, moreover, there is a competition between the As-Fe
superexchange and Fe-Fe exchange interaction. The combination of these
effects results in the highly anisotropic nearest neighbor exchange
interaction.

To clarify the relation between the structural transition and magnetic
property, we also perform calculation for low-temperature orthorhombic phase%
\cite{Low temp str}. Same with the high-temperature tetragonal structure,
for orthorhombic phase the S-AFM configuration is also lower in energy
comparing with other states. We reproduce that the ground state is the one
with the magnetic moments at the iron sites aligning antiparallel along the
longer \textit{a} axis. However, both the obtained magnetic moment (1.67 $%
\mu _{B}$) and the exchange interaction (J$_{1x}$=48.2, J$_{1y}$=-10.1, and J%
$_{3}$=21.1 meV) are almost the same as those in the high-temperature phase.
Moreover, we optimize lattice parameter and the internal atomic coordinate
for both stripe antiferromagnetic ordering (S-AFM) and checkboard
antiferromagnetic ordering (C-AFM). Our numerical results confirm that the
structure of Fe-pnictide is almost not depend on the magnetic configuration.
Therefore, exchange-striction effect, which had been used to explain the
uncentrosymmetric structural distortion and the associated multiferroics\cite%
{exchange striction}, cannot be used to explain the orthorhombic-tetragonal
transition. 

In summary, based on a combination of linear response theory and constrained
orbital hybridization approach, we study the mechanism of magnetic exchange
interaction of iron-based superconductor. Our results unambiguously identify
that the magnetic exchange process happens in the FeAs layer, and the highly
anisotropic feature of exchange interaction is not related to the orbital
polarization. The magnetism is at least partially itinerant, which results
in the anisotropic exchange interaction. While, the next nearest neighbor
interaction J$_{2}$ is mainly contributed by the As-bridged superexchange,
Fe-Fe exchange interaction has considerable effect on the nearest neighbor
exchange interaction J$_{1x}$ and J$_{1y}$.

The work was supported by National Key Project for Basic Research of China
(Grant No. 2011CB922101, and 2010CB923404), NSFC under Grant No. 10974082,
and 11174124.

\end{document}